\documentclass[12pt]{article}
\usepackage{amssymb}
\newcommand{\be}{\begin{equation}}
\newcommand{\bea}{\begin{eqnarray}}
\newcommand{\eea}{\end{eqnarray}}
\newcommand{\ba}{\begin{array}}
\newcommand{\ea}{\end{array}}
\newcommand{\ee}{\end{equation}}

\expandafter\ifx\csname mathbbm\endcsname\relax

\else Kim

\fi
\def\appendix{{\newpage\section*{Appendix}}\let\appendix\section%
        {\setcounter{section}{0}
        \gdef\thesection{\Alph{section}}}\section}

\begin{document}

\begin{titlepage}
\hfill
\vbox{
    \halign{#\hfil         \cr
           CERN-TH/2001-160 \cr
           hep-th/0106187  \cr
           } 
      }  
\vspace*{30mm}
\begin{center}
{\Large {\bf  String Field Theory and  the Fuzzy Sphere}\\} 

\vspace*{15mm}
\vspace*{1mm}
Harald Ita and Yaron Oz 

\vspace*{1cm} 

{\it Theory Division, CERN \\
CH-1211, Geneva  23, Switzerland}\\

\vspace*{.5cm}
\end{center}

\begin{abstract}
We use boundary string field theory to study open string tachyon condensation
on a three-sphere closed string background.
We consider the closed string background described by $SU(2)_k$
WZW model in the limit of large $k$. We compute the exact tachyon potential
and analyse the decay modes.

\end{abstract}
\vskip 4cm

June 2001

\end{titlepage}

\newpage

\section{Introduction}

Non-BPS D-brane systems and the process of open string tachyon condensation
have been extensively studied in recent years \cite{Sen:1999mg}.
In one line of research,
it has been shown that by using boundary string field theory (BSFT)
\cite{Witten:1992qy,Witten:1993cr,Shatashvili:1993ps,Shatashvili:1993kk}
the study of open string tachyon condensation
simplifies considerably \cite{Gerasimov:2000zp,Kutasov:2000qp,Kutasov:2000aq}.
In this case the system has only one field, the tachyon, and
one can compute exact properties such as profiles and tensions of lower
dimensional branes.
So far, boundary string field theory has been applied in the cases of a
flat target space. 
The inclusion of a nonzero closed string $B$-field in the BSFT framework
has been analysed
in \cite{Cornalba:2001ad,Okuyama:2001ch,Nemeschansky:2000iw}, and 
the inclusion of
the open string gauge fields has been studied in 
\cite{Andreev:2001yn,Kraus:2001nj,Takayanagi:2001rz}.

It is clearly of interest to extend the available methods
to  consider curved closed string backgrounds.
This is of particular interest since it can touch on the nature of the
background independence of the boundary string field theory formalism.

In this letter we will make a step into this direction.
We will consider as the closed string background the three-sphere $S^3$.
Closed strings on $S^3$ are described by $SU(2)_k$ WZW model. 
Tachyon condensation in this background has been discussed in 
\cite{Hikida:2001cp}.
We will be interested in the boundary string field theory description
of D-brane systems wrapping 2-cycles in $S^3$.  Exact analysis of these systems
when the level $k$ is finite is technically difficult.
However, in the limit of large $k$ the system simplifies considerably,
and allows an exact analysis.
We will compute the exact tachyon effective action up to two derivatives
in the tachyon field, and in particular the exact tachyon potential.
We will then use the results to
study the possible decay products.

The letter is organised as follows.
In section 2 we briefly review D-branes in $SU(2)_k$ WZW model and the
structure of their function algebra in the large $k$ limit.
In section 3 we will 
consider the bosonic boundary string field theory approach to D-branes
wrapping $S^2 \subset S^3$. We will perform exact computations
in the large $k$ limit and will obtain the exact tachyon
potential. We will then discuss the decay products. In
section 4 we will consider the
supersymmetric string case.
The relevant systems are D-branes and anti D-branes wrapping the same
2-cycle or different ones, and systems of non-BPS D-branes obtained from
the branes-antibranes systems by projection.  
We will compute the exact tachyon potential in all these cases and
analyse the decay products.

\section{$SU(2)$ WZW and the fuzzy sphere}

Consider closed strings propagating
on a three-sphere $S^3$  with radius $R_{S^3} = \sqrt{\alpha'k}$,
and a nonzero NSNS 3-form  $H$ field proportional to the volume form $\omega_{S^3}$.
The metric on $S^3$ is given by
\begin{eqnarray}
  \label{eq:S3geometry1}
  ds^{2}_{S_{3}}&=&\alpha' k\left(d\theta^2+\sin^2{\theta}
(d\psi^2+\sin^2{\psi}d\phi^2)\right).
\end{eqnarray}
This system is described by an
 $SU(2)_k$ WZW model.

Symmetry preserving D-branes\footnote{We will concentrate on the branes related to conjugacy 
classes with trivial auto-morphism which are centred around a fixed point 'the' pole 
of the $S^3$.} on $S^3$ are characterised
by  boundary conditions on the currents 
\be
J(z) = \bar{J}(\bar{z})~~~at~~~z=\bar{z} \ ,
\ee
where 
\begin{eqnarray}
  \label{eq:conscurr}
 J(z)=-k\partial_zgg^{-1},\quad \bar{J}(\bar{z})=kg^{-1}\partial_{\bar{z}}g \ .
\end{eqnarray}
Solutions to these boundary conditions can be labelled by an index $\alpha=0,1,...,k$,
with each 
D-brane having a world volume being an $SU(2)$ conjugacy class. 
Geometrically these 
conjugacy classes are two-spheres $S^2 \subset S^3$ specified by an angle
$\theta$ given by
\begin{eqnarray}
  \label{eq:S3geometry}
  \theta_{\alpha}=\frac{\pi \alpha}{k},\quad 0\le \alpha\le k \ .
\end{eqnarray}

For a D-brane with label $\alpha$ the open string
vertex operators are labeled by $V[Y_m^J]$ $J=0,1...,min(\alpha,k-\alpha), |m| \leq J$ 
where $m$ is an integer.
The OPE of the vertex operators $V[Y^J_m]$ reads \cite{Alekseev:1999bs,Runkel:1999pm}
\begin{eqnarray}
  \label{eq:vertexalg}
  V[Y^I_i](x_1)V[Y^J_j](x_2)\sim\sum_{K,k}(x_1-x_2)^{h_K-h_I-h_J}\left[\begin{array}{lll}I&J&K\\i&j&k\end{array}\right]c_{IJK}^{k,\alpha}\;V[Y^K_k](x_2) \ , 
\end{eqnarray}
where $h_I=\frac{I(I+1)}{k+2}$ is the conformal dimension of  $V[Y^I_i](x_1)$ and
 $[:::]$ are 
the Clebsch-Gordan coefficients of $SU(2)$. The structure constants 
$c_{IJK}^{k,\alpha}$ are given by the q-deformed $6J$ symbols of $SU(2)$ 
\cite{Alvarez-Gaume:1989vr},
\begin{eqnarray}
  \label{eq:qdef6j}
 c_{IJK}^{k,\alpha}=\left\{\begin{array}{lll}I&J&K\\\alpha/2&\alpha/2&\alpha/2\end{array}\right\}_q,\quad q=e^{\frac{2\pi i}{k+2}} \ .
\end{eqnarray}
 
We will consider the system in the limit of large $k$.
In this case
the q-deformation 
parameter goes to one and
the structure constants $c_{IJK}^{k,\alpha}$ become the ordinary $6J$ symbols.
We expand around small conformal weights, such that the OPE (\ref{eq:vertexalg}) 
depends on the insertion points only through the order of the operators
\begin{eqnarray}
  \label{eq:limit2}
  (x_1-x_2)^{h_K-h_I-h_J}\rightarrow 1,\;\;\; \mbox{for}\;\;\;\frac{I(I+1)}{k+2},\;\frac{J(J+1)}{k+2},...\ll1 \ .
\end{eqnarray}

To leading order the OPE of the vertex operators then reads
\begin{eqnarray}
  \label{eq:simpleOPE}
 V[Y^I_i]\cdot V[Y^J_j]=\sum_{K,k}\left[\begin{array}{lll}I&J&K\\i&j&k\end{array}\right]c_{IJK}^{k,\alpha}\;V[Y^K_k] \ .
\end{eqnarray}
The OPE (\ref{eq:simpleOPE}) 
for a brane $\alpha$ is isomorphic 
to the matrix algebra $M_{\alpha+1}({\mathbb C})$  \cite{Alekseev:1999bs}
\begin{eqnarray}
  \label{opeisom}
  V[Y^I_i]\sim Y_{Ii},\;\;\;V[Y^I_i]\cdot V[Y^J_j]\sim Y_{Ii}*Y_{Jj} \ ,
\end{eqnarray}
where $*$ is the ordinary matrix product (see appendix A for a detailed discussion).
This simplification can be used
in order to calculate the expectation values of products of vertex operators via
\begin{eqnarray}
  \label{eq:vertprod}
  \langle V[A](x_1)V[B](x_2)...V[C](x_n)\rangle
\sim \;tr(A*B*...*C) \ , 
\end{eqnarray}
where $V[A] = \sum_{Jj}A_{Jj}V[Y_j^J]$.

Note that since  
for a stack of N branes on the same world volume 
we have to add the  Chan 
Paton matrices $\lambda$,
the effective matrix algebra is enlarged from $M_{\alpha+1}({\mathbb C})$ 
to $M_{N(\alpha+1)}({\mathbb C})$.

\section{The bosonic BSFT}

In this section we will consider the boundary string field theory approach to D-branes
wrapping $S^2 \subset S^3$ in the large $k$ limit.
We will consider the bosonic case,
compute the exact tachyon potential and analyse the decay modes.
In general the D-branes may have extra world volume coordinates in $M^6$.
This can be taken into account simply by considering matrix valued functions
on these coordinates. In the following we will ignore dependence on $M^6$.

\subsection{The tachyon action}

Consider the two-dimensional action
\begin{eqnarray}
  \label{eq:openaction}
  S=S_0+\int_{\partial D}\!\!\!\!d\tau{\cal V}\ ,
\end{eqnarray}
where $S_0$ denotes an open plus closed
 conformally invariant background (the bulk action), and ${\cal V}$ is 
a general boundary perturbation. In our case $S_0$ is a  WZW action,
while ${\cal V}$ is 
\begin{eqnarray}
  \label{eq:pertbos}
  {\cal V}=T(g(z,\bar{z}))\vert_{z=\bar{z}}=\sum_{I,i} T^I_{i}V[Y_{i}^{I}](z,\bar{z})\vert_{z=\bar{z}} \ . 
\end{eqnarray}
${\partial D}$ denotes the boundary of the disk $D$.
Following \cite{Witten:1992qy,Witten:1993cr,
Shatashvili:1993ps,Shatashvili:1993kk} 
one 
constructs the partition function on the disk
\be
  \label{eq:partsum}
 Z(\lambda) = \langle \exp(\int_{\partial D}\!\!\!\!d\tau\sum_{Ii}T^I_{i}V[Y^I_i])\rangle_{S_0} \ .
\ee
The proposed space-time action ${\cal S}(T)$ is defined by 
\be
  \label{eq:boundentr}
  {\cal S}(T)=(\beta^{T}\partial_{T}+1)Z(T) \ ,
\ee
where $\beta^{T}$ are the $\beta$-functions of the couplings $T_{i}^{I}$.

We will work in the large $k$ limit described in the previous section.
The $\beta$-function for (the SU(2) part of) the tachyon reads 
\be
  \label{eq:tachbet}
  \beta_T = -T+\alpha'G^{ab}L_aL_bT + O(T^2) \ ,
\ee
where $L_a, a=1,2,3$ are the rescaled angular momentum operators
\be 
  L_a={\bf L}_a/\sqrt{2\alpha'}, \quad {\bf L}_a=[Y^1_a,.] \ ,
\ee
and
\be
G^{ab}=\left(\frac{2}{k}+O(1/k)^2\right)\delta^{ab} \ .
\ee

The string partition function (\ref{eq:partsum})
to linear order in $\alpha'$ reads
\begin{eqnarray}
  \label{eq:quadpartfunc}
  Z(T)=C_{\alpha}\;tr\left[e^{-T}(1-a_1\alpha'L_aL^a)T\right] \ ,
\end{eqnarray}
where $\alpha$ labels the $S^2$ conjugacy classes on the $S^3$ and $tr$ is taken
on the $(\alpha+1)\times(\alpha+1)$ matrices.
The normalisation factor 
\be
C_{\alpha} = \frac{4\pi\alpha'k\;sin(\frac{\pi \alpha}{k})}{\alpha} \ ,
\ee
is chosen such that for $T=0$ one gets the mass of the brane. 
The mass is given by the noncommutative brane tension.
The usual string coupling $g_s$ is replaced by the non commutative string 
coupling $G_s=\frac{g_s}{\sin{\frac{\pi \alpha}{k}}}$ along the lines of 
\cite{Hashimoto:2001xy} times the volume of the brane. 
$a_1$ is a numerical constant that will be fixed shortly.

The boundary string field theory
relates the string partition function to the space-time action (\ref{eq:boundentr})
\begin{eqnarray}
  \label{eq:effectact}
  {\cal S}(T)&=&(1+\beta_T\frac{\delta}{\delta T})Z(T)\\
      &=&C_{\alpha}tr\left[e^{-T}(1+T-\alpha'(1-a_1)(L_aL^a)T\right] \ .
\end{eqnarray}

Let us  show that 
by using appropriate field redefinitions the tachyon actions, 
to all orders in $T$ and to second order in ``derivatives'' $L_{a}$,
can thus be recast in the form
\begin{eqnarray}
  \label{eq:taac}
  {\cal S}(T) = C_{\alpha} tr\left[e^{-T}\left(L_aTL^{a}T+1+T\right)\right].
\end{eqnarray}

We start with an action for the tachyon of the form
\begin{eqnarray}
  \label{eq:orbitaction}
  {\cal S}(T)\sim tr\left[e^{-T}\left(a_1(T)L_aTa_{2}(T)L^{a}T+1+T\right)\right] \ ,
\end{eqnarray}
where $a_i(T)$ are some polynomials in $T$.
We can use a field redefinition
\begin{eqnarray}
  \label{eq:fieldredefinitions}
  T\rightarrow T-b_1(T)L_aTb_{2}(T)L^{a}T \ ,
\end{eqnarray}
where $b_i(T)$ are some polynomials in $T$,
which generates, to leading order in $L_aL^a,$ an orbit of actions
\begin{eqnarray}
  \label{eq:orofac}
  {\cal S}_{orbit}(T)\sim tr\left[e^{-T}\left(a_1(T)L_aTa_{2}(T)L^{a}T+Tb_1(T)L_aTb_{2}(T)L^{a}T+1+T\right)\right] \ .
\end{eqnarray}
These orbits of actions thus differ by a kinetic term of the form 
\be
{\cal S}_{kin} = C\;e^{-T}L_aTL^aT \ ,
\ee
where $C$ is a constant, which 
cannot be generated by field redefinitions.

Thus, the most general tachyon action is
\begin{eqnarray}
  \label{eq:orbitaction1}
  {\cal S}(T)\sim 
tr\left[e^{-T}\left(\sum_n a_{1,n}(T)L_aTa_{2,n}(T)L^{a}T+1+T\right)\right] \ .
\end{eqnarray}
and as we have seen it can be reached by field redefinitions
from the action with a constant in front the kinetic term.
To fix the constant we can use 
the  consistency of the field equations derived from ${\cal S}(T)$ and the $\beta$-function 
\cite{Gerasimov:2000zp}
\be
  \label{eq:consiteffactbeta}
  \beta_T\sim\frac{\delta {\cal S}(T)}{\delta T} \ .
\ee
We get
\begin{eqnarray}
  \label{eq:copi}
  C=\sum_n a_{1,n}(0)a_{2,n}(0)=1 \ ,
\end{eqnarray}
and arrive at (\ref{eq:taac}).

\subsection{Decay modes}

We have seen above that for a D-brane wrapping the $\alpha$ conjugacy class,
the tachyon potential is given by 
\be
V(T) = C_{\alpha}tr[e^{-T}\left(1+T\right)] \ .
\label{potential}
\ee
For a single D-brane,
$T$ is a hermitian $(\alpha+1)\times(\alpha+1)$ matrix.
For $N$ D-branes we have to increase the matrix size to  $N(\alpha+1)\times N(\alpha+1)$.
The potential (\ref{potential})
is the exact (string tree level) tachyon potential in the large $k$ limit.
It 
has the same form as for a flat bosonic D-brane and the flat bosonic D-brane with a constant
$B$-field
in the non-commutative limit. 
What differs is the function algebra, which in our case
is simply a matrix algebra. 
In the following we discuss the possible decay modes of the system.
For this analysis we can neglect the kinetic term and analyse
the potential.

Minima of the tachyon potential satisfy
\begin{eqnarray}
  \label{eq:bostachpot}
  V'(T)=-T e^{-T}=0 \ .
\label{ex}
\end{eqnarray}
In order to analyse (\ref{ex}) we can diagonalise the tachyon.
Solutions of (\ref{ex}) are matrices with zero and infinite eigenvalues.
The top of the tachyon potential is when $T$ is the zero matrix.
The absolute minimum of the tachyon potential 
is reached when the tachyon's eigenvalues are all infinite
\begin{eqnarray}
  \label{eq:comlcond1}
  T=\lambda\; diag\{t_1,t_2,....t_{\alpha+1}\},\quad \lambda\rightarrow\infty, \quad t_i>0 \ .
\end{eqnarray}
It corresponds to reaching the closed string vacuum.

Intermediate decay products correspond to tachyon configurations 
where not all eigenvalues are infinite
\begin{eqnarray}
  \label{eq:metadecay}
  T=\lambda\; diag\{t_1,..,t_k,0,..,0\},\quad \lambda\rightarrow\infty, \quad t_i>0 \ .
\end{eqnarray}
Such a configuration corresponds to the decay of the $\alpha$ D-brane
to the $\alpha-k$ D-brane. The number of zeros
is also the number of D0-branes from which the spherical D2-brane
is built.
In particular
the trivial tachyon $T=0$ describes the $\alpha$
D2-brane  which is made of $\alpha+1$ D0-branes.

In the language of perturbations (\ref{eq:pertbos}) the analysis of decay modes is simple.
Since we work in the large $k$ limit we associate for a given tachyon perturbation an
$(\alpha+1)\times(\alpha+1)$ matrix,
using the results in section 2 and the details in the appendix. 
We then diagonalise this matrix and analyse its eigenvalues as above.
This provides us
with the information on the decay mode associated with the perturbation
and a simple picture of the endpoint of the two-dimensional
RG flow.
Let us illustrate this with
a simple example.

We consider
a D2-brane wrapping the $\alpha=1$ 2-sphere. 
The tachyon field is a hermitian $2\times2$-matrix $T$, which we can
expand in terms of the matrix representation of the $\alpha=1$
matrices  $\{Y_{j}^J\}_{ab}$ given by (\ref{eq:matrices}). 
When $T$ has two positive eigenvalues the system condenses to the vacuum.
This happens, for instance, if
$T=iY^0_0$.  
To get an $\alpha=0$ brane at the endpoint of the condensation we can take
a tachyon configuration of the form
$T=Y^1_1+Y^1_{-1}-i\sqrt{2}Y^0_0$ which has eigenvalues zero and one.

We note that this simple picture is really a feature of the large $k$ limit.
For finite $k$, such perturbations which are typically not free, are much harder
to analyse. In particular, we cannot simply 
look for the vanishing locus of the
tachyon profile expressed via the spherical harmonics in order to analyse the decay
product, as done for the free perturbations used in the flat target space case.
This can be seen from the above example. The spherical harmonic $iY^1_0(\theta,\phi)$ has zeros at the poles $\psi=0,\pi$. Thus one might suspect the $\alpha=1$ brane to decay into two separated $\alpha=0$ branes on the poles. However, the matrix $iY^1_0$ has nonzero eigenvalues $\pm1/\sqrt{2}$.

It seems plausible to assume that for a given tachyon perturbation, the
endpoint of the two-dimensional
RG flow for finite $k$ 
will not differ from  the large $k$ one.
In such a case, while the finite $k$ RG-flow is complicated
to analyse, at least the end-points of the 
flow have a simple picture as described by the large
large $k$ limit above.

\section{Supersymmetric BSFT}
In this section we will consider the supersymmetric
boundary string field theory approach to $D\bar{D}$-branes and non-BPS branes
wrapping $S^2 \subset S^3$ in the large $k$ limit.

\subsection{The tachyon action}

The boundary vertex operators for $N_{\alpha}$ branes and $N_{\beta}$ antibranes can be constructed from \cite{Fuchs:1987ew,Fuchs:1989gm},
\begin{eqnarray}
  \label{eq:bouncoup}
  M(A_{\alpha},A_{\beta},T)&\!\!=&\!\!\left(\begin{array}{cc} -iA_{\alpha}&T\\T^{\dagger}&-iA_{\beta}\end{array}\right),
\end{eqnarray}
with,
\begin{eqnarray}
  \label{eq:susyvert}
A_i&\!\!=&\!\!\psi_a V[A^a_i]+\theta\sqrt{{2}/{\alpha'}}\left[:j_a V[A^a_i]:-\alpha':\psi_a\psi_b:V[L^bA^a_i]\right],\quad i=\alpha,\beta\\
T&\!\!=&\!\!(V[T]-\theta\sqrt{2\alpha'}\left[\;\psi_aV[L^aT]\right] \ .
\end{eqnarray}
where $\psi_a$ and $j_a$ form the supercurrent 
$J_a(x)=i\sqrt{2}(\psi_a(x)-\sqrt{2/\alpha'}\theta j_a(x))$. $A^a_i$ are products of $Y^J_j$s and 
Chan Paton matrices of dimensions $N_i \times N_i$, such that $V[A^a_i]^{\dagger}=V[A^a_i]$. 
The tachyon fields T are complex ($T=\frac{1}{2}(T_1+iT_2)$) 
products of $Y^J_j$s and Chan Paton matrices of dimensions $N_{\alpha} \times N_{\beta}$. 
The superconnection structure appears in(\ref{eq:bouncoup}), as in 
\cite{Witten:1998cd,Kennedy:1999nn,Kraus:2001nj,Takayanagi:2001rz,Alishahiha:2001du}.

The operators $L_aT$ act as the generators of the rotation group on the respective branes. For the tachyon field they are defined as $\sqrt{2\alpha'}L_aT=((Y_{\alpha})^1_a\otimes {1}_{\alpha\times\alpha})T-T((Y_{\beta})^1_a\otimes {1}_{\beta\times\beta})$. This is natural as the tachyons transform in the bifundamental representation of the rotation groups on the respective branes $\alpha,\beta$. 
Note that
the Chan Paton factors are not changed under the rotation.  
The normalisation is such that the OPEs read 
\begin{eqnarray}
  \label{eq:CFTscal}
  j^a(z_1)j^b(z_2)&=&\frac{\alpha'}{2}\frac{G^{ab}}{(z_1-z_2)^2}+\alpha'\frac{if^{ab}_{\;\;\;c}j^c(z_2)}{z_1-z_2} \ ,\\
  \psi^a(z_1)\psi^b(z_2)&=&\frac{G^{ab}}{z_1-z_2}, \quad f^{ab}_{\;\;\;c}=\frac{1}{\sqrt{\alpha'}}\frac{2}{k}\varepsilon^{ab}_{c} \ .
\end{eqnarray}

For later use we introduce $M_0$ and $M_1$ for the $\theta$ independent and the $\theta$ dependent part of $M=M_0+\theta M_1$. $M_0$ ($M_1$) is related to the picture $(-1)$ (picture $(0)$) vertex operators of the sum of the tachyon and the gauge field. 

The world sheet action reads 
\begin{eqnarray}
  \label{eq:boundcoup}
 S&=&S_0+S_{pert} \ ,\\
 S_{pert}&=&\int d\hat{\tau}\left[\hat{\bar{\eta}}^aD\hat{\eta}^a+\hat{\bar{\eta}}^aM_{ab}\hat{\eta}^b\right] \ ,
\end{eqnarray}
where the boundary fields $\hat{\bar{\eta}},\hat{\eta}$ have to be integrated over in the
 path integral. We use the notation $d\hat{\tau}=d\tau d\theta$, $D=\partial_{\theta}+\theta\partial_{\tau}$ and the fermionic superfields $\hat{\eta}^a=\eta^a+\theta\chi^a$.

The integration over the auxiliary fields $\chi$ rearranges the boundary action, such that one finds the path ordered product of $M_0^2+M_1$,
\begin{eqnarray}
  \label{eq:effpert}
 S&=&S_0+S_{pert}' \ ,\\
 S_{pert}'&=&\left(\int d\tau\left[\bar{\eta}\partial_{\tau}\eta+\bar{\eta}(M_0^2+M_1)\eta\right)\right] \ .
\end{eqnarray}

For zero gauge fields $A_1,A_2$ one finds 
\begin{eqnarray}
  \label{eq:effM}
  M_0^2+M_1=\left(\begin{array}{cc}V[T]V[T]^{\dagger}&\;-\sqrt{2\alpha'}\;\psi_aV[L^aT]\\-\sqrt{2\alpha'}\;\psi_aV[L^aT]^{\dagger}&V[T]^{\dagger}V[T] \end{array}\right) \ .
\end{eqnarray}

From this (\ref{eq:effM}) the tachyon potential, which is the leading order term in the $1/k$ expansion, can be read. First order corrections have two origins: corrections from the OPE (\ref{eq:vertexalg}) and contributions from the off diagonal entries in (\ref{eq:effM}).
\begin{eqnarray}
  \label{eq:renpart}
  Z(T,T^{\dagger})&=&C_{\alpha}tr_{\alpha}\left[e^{-c_1TT^{\dagger}-c_2LT(LT)^{\dagger}}\right]\\
  &&\quad+C_{\beta}tr_{\beta}\left[e^{-c_1T^{\dagger}T-c_2(LT)^{\dagger}LT}\right]+
O(\frac{1}{k^2}) \ .
\end{eqnarray}
The kinetic terms are understood to be ordered in a  symmetric way. The constants $c_1,c_2$ will be fixed below.
The BSFT action ${\cal S}$ of the super string is conjectured to be \cite{Kutasov:2000aq}
(see also \cite{Marino:2001qc, Niarchos:2001si})  
\begin{eqnarray}
  \label{eq:partfunc}
  {\cal S}(T)=Z(T) \ .
\end{eqnarray}

Consistency with the $\beta$ function can be used to fix the coefficients $c_1=1/4, c_2=-\alpha'/2$. This is because to quadratic order in the tachyon the action has to reproduce the mass formula $-\alpha'm^2=J(J+1)/k-1/2$. The kinetic term gives
\begin{eqnarray}
  \label{eq:confweight}
 2\frac{\alpha'}{2}G^{ab}L_aL_bY^J_j=\frac{J(J+1)}{k}Y^J_j \ . 
\end{eqnarray}
The constant $c_2$ is fixed such that the quadratic term in the tachyon potential produces the $-\frac{1}{2}$ in the open string mass formula.

Finally, the tachyon action up to second order in $1/k$ reads
\begin{eqnarray}
  \label{eq:tachact}
  {\cal S}(T)_{D\bar{D}}&=&C_{\alpha}tr_{\alpha}\left[e^{-\frac{1}{4}TT^{\dagger}}\left(-\alpha'/2\;L_aT (L^aT)^{\dagger}+1\right)\right]\\&&\quad+C_{\beta}tr_{\beta}\left[e^{-\frac{1}{4}T^{\dagger}T}\left(-\alpha'/2\;(L_aT)^{\dagger} L^aT+1\right)\right] \ .
\end{eqnarray}

Next we will derive the tachyon action for a $N_{\alpha}$ non-BPS branes on a $\alpha$ sphere $S_2\subset S_3$. To this end one gauges  the $(-1)^{F_L}$ symmetry  of the system of $N_{\alpha}$ brane anti brane pairs \cite{Sen:1999mg}. This equates the gauge fields $A_{\alpha}=A_{\alpha'}$ and selects the tachyon field to be real $T=T^{\dagger}$.

The boundary perturbation simplifies to
\begin{eqnarray}
  \label{eq:bCCouncoupn}
  M(A,T)&=&\left(\begin{array}{cc} -iA&T\\T&-iA\end{array}\right) \ .
\end{eqnarray} 
Analogous steps as above lead to the tachyon action
\begin{eqnarray}
  \label{eq:tachactn}
  {\cal S}(T)_{non-BPS}=\sqrt{2}C_{\alpha}tr_{\alpha}\left[e^{-\frac{1}{4}TT}\left(-\alpha'/2\;L_aT (L^aT)+1\right)\right] \ .
\end{eqnarray}

\subsection{Decay modes}

We can 
distinguish three different systems:\\
(1) Coinciding
brane-antibrane system wrapping the same conjugacy class $\alpha$, which
we denote
by $D(\alpha)\bar{D}(\alpha)$. \\
(2) Brane wrapping  a conjugacy class $\alpha$ and antibrane wrapping a
different  conjugacy class $\beta$, which we denote
by $D(\alpha)\bar{D}(\beta)$.\\
(3) Non-BPS brane wrapping a  conjugacy class $\alpha$.\\
Consider the decay modes of these systems.

\vskip 0.5cm
{\bf $D(\alpha)\bar{D}(\alpha)$ system }\\

The tachyon energy of  the $D2\bar{D2}$ system wrapping the $\alpha$ 2-cycle is given by
\begin{eqnarray}
  \label{eq:tchpotsusy}
  E(\alpha,T,T^{\dagger})= 2C_{\alpha}tr\left(e^{-\frac{1}{4}TT^{\dagger}}\right) \ .
\end{eqnarray}
As in the bosonic case we are interested in the extrema of this potential, they fulfil the equations
\begin{eqnarray}
  \label{eq:extrsusy}
  T^{\dagger}e^{-\frac{1}{4}TT^{\dagger}}=0,\quad e^{-\frac{1}{4}TT^{\dagger}}T=0 \ .
\end{eqnarray}
The tachyon can be diagonalised by two unitary matrices $U,W$ with $UTW^{\dagger}$. 
The global minimum corresponds to tachyon perturbations with nonzero
eigenvalues that become infinite at the endpoint of the perturbation
\begin{eqnarray}
  \label{eq:comlcond}
  T=\lambda\; diag\{t_1,t_2,....t_{\alpha+1}\},\quad \lambda\rightarrow\infty, \quad t_i\neq0 \ .
\end{eqnarray}

Intermediate decay products of $D2\bar{D2}$ system wrapping a 2-cycle $\beta < \alpha$
correspond to the tachyon matrices with some zero eigenvalues. 
They are bi-unitary transformations of
\begin{eqnarray}
  \label{eq:smalld2}
  T=\lambda\; diag\{t_1,..,t_k,0,..,0\}_{\alpha+1},\quad \lambda\rightarrow\infty, \quad t_i\neq0 \ .
\end{eqnarray}
Tachyons of the form (\ref{eq:smalld2}) correspond to a perturbation from a  $D2\bar{D2}$
system wrapping a 2-cycle $\alpha$ to a  $D2\bar{D2}$ system wrapping
a 2-cycle $\beta = \alpha-k$.
One can also interpret the zeros of the tachyon matrix
as counting
 the number of $D0\bar{D0}$ constituent states of the condensed $D2\bar{D2}$ system. 
As usual the zero tachyon
corresponds to the top of the potential
with no condensation.

\vskip 0.5cm
{\bf $D(\alpha)\bar{D}(\beta)$ system }\\

We consider a brane wrapping a 2-cycle $\alpha$ and an antibrane wrapping a
2-cycle $\beta$.
Without loss of generality we assume that $\alpha>\beta$ and that the branes are concentric. 
The energy of the system in this case reads
\begin{eqnarray}
  \label{eq:mixedpot}
  E(\alpha,\beta,T,T^{\dagger})&=&C_{\alpha}tr_{\alpha}\left(e^{-\frac{1}{4}TT^{\dagger}}\right)+C_{\beta}tr_{\beta}\left(e^{-\frac{1}{4}T^{\dagger}T}\right) \ .
\end{eqnarray}
For extrema the conditions 
\begin{eqnarray}
  \label{eq:minimix}
  T^{\dagger}e^{-\frac{1}{4}TT^{\dagger}}=0 \ , \quad e^{-\frac{1}{4}TT^{\dagger}}T=0
\end{eqnarray}
have to hold. 

We can think of the tachyon $T$ and its conjugate $T^{\dagger}$
as maps $T: E_{\beta} \rightarrow E_{\alpha}$ and $T^{\dagger}: E_{\alpha} \rightarrow E_{\beta}$,
where
$E_{\alpha}$ and $E_{\beta}$ are the vector bundles on the $\alpha$ and $\beta$ 2-cycles
corresponding to the brane and antibrane respectively.
The relevant operators for the discussion are $TT^{\dagger}$ and $T^{\dagger}T$.
In our case, they are matrices of size $(\alpha+1)\times(\alpha+1)$ and 
$(\beta+1)\times(\beta+1)$.
Their zeros determine the number of $D$0-branes and ${\bar{D}}$0-branes constituents,
respectively. Thus, the index 
\be
Index(E,T) = dim~Ker~TT^{\dagger} - dim~Ker~T^{\dagger}T \ ,
\ee
counts the net $D$0-brane charge.

As to the decay modes, the analysis is as before. A  matrix $TT^{\dagger}$
of the form
\begin{eqnarray}
  \label{eq:smalld3}
 TT^{\dagger} =\lambda\; diag\{t_1,..,t_k,0,...,0\}_{\alpha+1},\quad \lambda\rightarrow\infty, 
\quad t_i\neq0 \ ,
\end{eqnarray}
corresponds to a perturbation that will reduce 
the 2-cycle $\alpha$ to 
a 2-cycle $\alpha-k$.
 A  $T^{\dagger}T$ matrix
of the form
\begin{eqnarray}
  \label{eq:smalld4}
 T^{\dagger}T =\lambda\; diag\{t_1,..,t_k,0,...,0\}_{\beta+1},\quad \lambda\rightarrow\infty, 
\quad t_i\neq0 \ ,
\end{eqnarray}
corresponds to a perturbation that will reduce 
the 2-cycle $\beta$ to 
a 2-cycle $\beta-k$.

\vskip 0.5cm
{\bf Non-BPS branes}\\

For a non-BPS $\alpha$ brane wrapping the $\alpha$ 2-cycle
the energy reads
\begin{eqnarray}
  \label{eq:nonpot}
  E(\alpha,T)&=&\sqrt{2}C_{\alpha}tr_{\alpha}\left(e^{-\frac{1}{4}TT}\right) \ ,
\end{eqnarray}
where $T=T^{\dagger}$.
The analysis of the decay modes is as before.
A tachyon of the form 
\begin{eqnarray}
  \label{eq:condpert}
 T=\lambda\; diag\{t_1,...t_k,0,...,0\}_{\alpha+1}  \ , 
\end{eqnarray}
initiates a  flow to a non-BPS branes wrapping the
$\alpha-k$ 2-cycle.

\vspace{5ex} {\bf Acknowledgement}:
We would like 
to thank S. Shatashvili for a valuable discussion. 

\nopagebreak
\begin{appendix}
{Representations Of The Fuzzy Algebra}

In this appendix we 
review some details concerning the
vertex operator algebra as  the matrix algebra of the fuzzy sphere
 \cite{Hoppe:1989gk}.
Recall the OPE of the vertex operators (\ref{eq:simpleOPE})
\begin{eqnarray}
  \label{eq:simpleOPE1}
 V[Y^I_i]\cdot V[Y^J_j]=\sum_{K,k}\left[\begin{array}{lll}I&J&K\\i&j&k\end{array}\right]c_{IJK}^{k,\alpha}\;V[Y^K_k] \ .
\end{eqnarray}
The symbols $[:::]$ are related to the  $3j$-symbols $(:::)$ by
\begin{eqnarray}
  \label{eq:clebthreej}
 \left[\begin{array}{lll}I&J&K\\i&j&k\end{array}\right]=(-1)^{k+1}\sqrt{2 I+1}\sqrt{2 J+1}\sqrt{2 K+1}\left(\begin{array}{rrr}I&J&K\\i&j&-k\end{array}\right).
\end{eqnarray}
The spherical harmonics $Y_{j}^J$ are represented on the $\alpha$ fuzzy sphere by the 
$(\alpha+1)\times(\alpha+1)$ matrix
\begin{eqnarray}
  \label{eq:matrreps}
 \{Y_{j}^J\}_{ab}&=&(-1)^{\alpha/2-j}\sqrt{2J+1}\left(\begin{array}{rrr}\alpha/2&J&\alpha/2\\-a&j&b\end{array}\right)\quad a,b=-\alpha/2,-\alpha/2+1,...,\alpha/2 \, \nonumber\\
 tr[Y_{j}^JY_{i}^I]&=&(-1)^{j+\alpha}\delta_{JI}\delta_{j,-i} \ .
\end{eqnarray}
For $\alpha=1$ the matrices $\{Y_{j}^J\}_{ab}$ are 
\begin{eqnarray}
  \label{eq:matrices}
  Y^0_0&=&\left(\begin{array}{cc}-i/\sqrt{2}&0\\0&-i/\sqrt{2}\end{array}\right),\\
  Y^1_1&=&\left(\begin{array}{cc}0&0\\-i&0\end{array}\right),\;Y^1_0=\left(\begin{array}{cc}i/\sqrt{2}&0\\0&-i/\sqrt{2}\end{array}\right),\;Y^1_{-1}=\left(\begin{array}{cc}0&i\\0&0\end{array}\right).
\end{eqnarray}
The most general tachyon configuration thus reads $T=\sum_{Jj}T^J_jY^J_j$. 
Its eigenvalues determine the endpoint of the respective RG flow, as discussed
in the paper.

\end{appendix}


\newpage

\begin{thebibliography}{99}


\bibitem{Sen:1999mg}
A.~Sen,
``{\it Non-BPS states and branes in string theory},''
hep-th/9904207.


\bibitem{Witten:1992qy}
E.~Witten,
``{\it On background independent open string field theory},''
Phys.\ Rev.\ D {\bf 46} (1992) 5467
[hep-th/9208027].

\bibitem{Witten:1993cr}
E.~Witten,
``{\it Some computations in background independent off-shell string theory},''
Phys.\ Rev.\ D {\bf 47} (1993) 3405
[hep-th/9210065].

\bibitem{Shatashvili:1993ps}
S.~L.~Shatashvili,
``{\it On the problems with background independence in string theory},''
hep-th/9311177.

\bibitem{Shatashvili:1993kk}
S.~L.~Shatashvili,
``{\it Comment on the background independent open string theory},''
Phys.\ Lett.\ B {\bf 311} (1993) 83
[hep-th/9303143].


\bibitem{Gerasimov:2000zp}
A.~A.~Gerasimov and S.~L.~Shatashvili,
``{\it On exact tachyon potential in open string field theory},''
JHEP {\bf 0010}, 034 (2000)
[hep-th/0009103].

\bibitem{Kutasov:2000qp}
D.~Kutasov, M.~Marino and G.~Moore,
``{\it Some exact results on tachyon condensation in string field theory},''
JHEP {\bf 0010}, 045 (2000)
[hep-th/0009148].

\bibitem{Kutasov:2000aq}
D.~Kutasov, M.~Marino and G.~Moore,
``{\it Remarks on tachyon condensation in superstring field theory},''
hep-th/0010108.


\bibitem{Cornalba:2001ad}
L.~Cornalba,
``{\it Tachyon condensation in large magnetic fields with background
  independent string field theory},''
Phys.\ Lett.\ B {\bf 504}, 55 (2001)
[hep-th/0010021].



\bibitem{Okuyama:2001ch}
K.~Okuyama,
``{\it Noncommutative tachyon from background independent open string field  theory},''
Phys.\ Lett.\ B {\bf 499}, 167 (2001)
[hep-th/0010028].


\bibitem{Nemeschansky:2000iw}
D.~Nemeschansky and V.~Yasnov,
``{\it Background independent open string field theory and constant B-field},''
hep-th/0011108.

\bibitem{Andreev:2001yn}
O.~Andreev,
``{\it
Some computations of partition functions and tachyon potentials in  background independent off-shell string theory},''
Nucl.\ Phys.\ B {\bf 598}, 151 (2001)
[hep-th/0010218].


\bibitem{Kraus:2001nj}
P.~Kraus and F.~Larsen,
``{\it Boundary string field theory of the D D-bar system},''
Phys.\ Rev.\ D {\bf 63}, 106004 (2001)
[hep-th/0012198].

\bibitem{Takayanagi:2001rz}
T.~Takayanagi, S.~Terashima and T.~Uesugi,
``{\it Brane-antibrane action from boundary string field theory},''
JHEP {\bf 0103}, 019 (2001)
[hep-th/0012210].


\bibitem{Hikida:2001cp}
Y.~Hikida, M.~Nozaki and T.~Takayanagi,
``{\it Tachyon condensation on fuzzy sphere and noncommutative solitons},''
Nucl.\ Phys.\ B {\bf 595}, 319 (2001)
[hep-th/0008023].



\bibitem{Alekseev:1999bs}
A.~Y.~Alekseev, A.~Recknagel and V.~Schomerus,
``{\it Non-commutative world-volume geometries: Branes on SU(2) and fuzzy  spheres},''
JHEP{\bf 9909} (1999) 023
[hep-th/9908040].

\bibitem{Runkel:1999pm}
I.~Runkel,
``{\it Boundary structure constants for the A-series Virasoro minimal models},''
Nucl.\ Phys.\ B {\bf 549} (1999) 563
[hep-th/9811178].


\bibitem{Alvarez-Gaume:1989vr}
L.~Alvarez-Gaume, C.~Gomez and G.~Sierra,
``{\it Quantum Group Interpretation Of Some Conformal Field Theories},''
Phys.\ Lett.\ B {\bf 220} (1989) 142.


\bibitem{Hashimoto:2001xy}
K.~Hashimoto and K.~Krasnov,
``{\it D-brane solutions in non-commutative gauge theory on fuzzy sphere},''
hep-th/0101145.

\bibitem{Fuchs:1987ew}
J.~Fuchs,
``{\it Superconformal Ward Identities And The WZW Model},''
Nucl.\ Phys.\ B {\bf 286} (1987) 455.

\bibitem{Fuchs:1989gm}
J.~Fuchs,
``{\it More On The Super WZW Theory},''
Nucl.\ Phys.\ B {\bf 318} (1989) 631.

\bibitem{Witten:1998cd}
E.~Witten,
``{\it D-branes and K-theory},''
JHEP {\bf 9812}, 019 (1998)
[hep-th/9810188].


\bibitem{Kennedy:1999nn}
C.~Kennedy and A.~Wilkins,
``{\it Ramond-Ramond couplings on brane-antibrane systems},''
Phys.\ Lett.\ B {\bf 464}, 206 (1999)
[hep-th/9905195].

\bibitem{Alishahiha:2001du}
M.~Alishahiha, H.~Ita and Y.~Oz,
``{\it On superconnections and the tachyon effective action},''
Phys.\ Lett.\ B {\bf 503}, 181 (2001)
[hep-th/0012222].

\bibitem{Marino:2001qc}
M.~Marino,
``{\it On the BV formulation of boundary superstring field theory,}''
JHEP {\bf 0106}, 059 (2001)
[hep-th/0103089].


\bibitem{Niarchos:2001si}
V.~Niarchos and N.~Prezas,
``{\it Boundary superstring field theory,}''
hep-th/0103102.

\bibitem{Hoppe:1989gk}
J.~Hoppe,
``{\it Diffeomorphism Groups, Quantization And SU(Infinity)},''
Int.\ J.\ Mod.\ Phys.\ A {\bf 4} (1989) 5235.

\end{thebibliography}
\end{document}